\documentclass[prl,twocolumn,aps,english,showpacs,longbibliography,superscriptaddress]{revtex4-1}
\usepackage{natbib}

\usepackage{graphicx}
\usepackage{xcolor}
\definecolor{deeppink}{rgb}{1.0, 0.08, 0.58}
\usepackage{bm}
\usepackage{mathptmx}

\usepackage[english]{babel}

\usepackage[colorinlistoftodos]{todonotes}
\usepackage[colorlinks=true, allcolors=blue]{hyperref}

\begin{document}

\title{Role of anti-phase boundaries in the formation of magnetic domains in magnetite thin films}

\author{Roberto~Moreno}
\affiliation{Department of Physics, University of York, Heslington, York, YO10 5DD, UK}
\affiliation{Earth and Planetary Science, School of Geosciences, University of Edinburgh, Edinburgh EH9 3FE, UK}
\author{Sarah~Jenkins}
\affiliation{Department of Physics, University of York, Heslington, York, YO10 5DD, UK}
\author{Vlado~K.~Lazarov}
\affiliation{Department of Physics, University of York, Heslington, York, YO10 5DD, UK}
\author{Richard~F.~L.~Evans}
\affiliation{Department of Physics, University of York, Heslington, York, YO10 5DD, UK}

\begin{abstract}
Anti-phase boundaries (APBs) are structural defects which have been shown to be responsible for the anomalous magnetic behaviour observed in different nanostructures. Understanding their properties is crucial in order to use them to tune the  properties of magnetic  materials by growing APBs in a controlled way since their density strongly depends on the synthesis method. With this aim, in this work we investigate their influence on magnetite (Fe$_3$O$_4$) thin films by considering an atomistic spin model, focusing our study on the role that the exchange interactions play across the APB interface. We conclude that the main atypical features reported experimentally in this material are well described by the model we propose here, confirming the new exchange interactions created in the APB as the responsible for this deviation from bulk properties.
\end{abstract}
\pagebreak
        
\maketitle
\section{Introduction}
Anti-phase boundaries (APBs) are stacking defects appearing in crystalline nanosystems, created by a displacement between atomic planes. They have been observed in many different materials  such as MnAl \cite{doi:10.1063/1.1782138}, GaAs \cite{pub1058136097} or magnetite (Fe$_3$O$_4$) \cite{pub1060580820}, as well as in different types of nanostructures, e.g,  thin films \cite{pub1060580820} or nanoparticles \cite{Nedelkoski2017}. The type and number of APBs appearing in nanostructures strongly depends on the synthesis method used to grow them. Thus, APBs have been suggested to be responsible of the different magnetic properties observed in samples made of the same material, with the same nanostructure, but created with different techniques, e.g, the saturation magnetization in magnetite nanoparticles \cite{Nedelkoski2017} or the  magnetic anisotropy in magnetite thin films \cite{PhysRevLett.79.5162,VANDERHEIJDEN199871,C6TC02152B}. In fact, recent works on magnetite thin films have given confidence on this supposition, showing that when the nanostructure is grown achieving a low density of APBs \cite{C6TC02152B,C7NR07143D} or reducing their strength by applying an electric field during the synthesis \cite{pub1104047228}, the magnetic properties not only begin to match between all the samples, but they start to be similar to the bulk case. 


To explain why APBs could strongly influence the magnetic properties in a system it is necessary to take into account that the number of interacting atoms across the APB interface and their corresponding distances might be modified due to the structural mismatch. Thus, a new set of exchange interactions ($J^{APB}_{ij}$) should appear locally in the APB, while the number of bulk exchange interactions ($J^{Bulk}_{ij}$) might be changed \cite{PhysRevLett.79.5162}. For the specific case of magnetic oxides, e.g, magnetite, for which the exchange interactions are mediated via the oxygen atoms, the angle that forms the two interacting magnetic atoms and the oxygen one might be also modified in the APB, consequently, the new set of ($J^{APB}_{ij}$) could have both different strength or sign than the ($J^{Bulk}_{ij}$).  Therefore, if $J^{APB}_{ij}$ dominates rather than $J^{Bulk}_{ij}$, the system could have a completely different magnetic behavior. Indeed, considering $J^{APB}_{ij}$ as antiferromagnetic instead of ferromagnetic has been useful to explain pinning effects in MnAl thin films \cite{PhysRevB.96.224411} or to qualitatively explain the magnetization reduction in magnetite nanoparticles  \cite{Nedelkoski2017}.  

For the specific case of magnetite thin films, which is one of the most promising materials for the next generation of spintronic devices due it is half metallic character \cite{pub1060558938,pub1063107957} and high Curie temperature ($T_C$ = 860 K \cite{PhysRev.186.577}), APBs seem to play a fundamental role on the magnetic properties of the system \cite{PhysRevLett.79.5162}. Firstly, both out of plane anisotropy \cite{PhysRevLett.79.5162} and fourfold in plane anisotropy \cite{C6TC02152B} have been observed for different samples. Even for the case of the fourfold anisotropy, different works report different easy axes \cite{VANDERHEIJDEN199871,C6TC02152B} suggesting APBs to be the source of this discrepancy. Secondly, it has been reported that the magnetization is not saturated under the effect of high magnetic fields of 70 kOe \cite{PhysRevLett.79.5162}, which means that exchange interactions are most likely to be responsible of this behaviour. Additionally, a recent work suggested that the strength of APB exchange interactions is reduced by applying electric fields during the synthesis process, as a consequence showing saturation fields of $\mu_0 H = 150$ mT \cite{pub1104047228} which is similar to bulk results. Finally, it has been demonstrated that when the samples are grown with low density of APBs, the number of magnetic domains observed is smaller than in samples with lots of APBs, pointing to them as a source of magnetic domains \cite{C7NR07143D}. 

As APBs are atomic-scale defects and their importance seems to come from the exchange interactions, we decided to study their influence on the magnetic properties using an atomistic spin model. Specifically, we use the open source VAMPIRE \cite{0953-8984-26-10-103202} software package to investigate their influence  in magnetite thin films, focusing on explaining the anomalous saturation magnetization as well as the reason of why samples with more APBs have more magnetic domains, leaving the fact of observing different magnetic anisotropies for future work. 


With this aim, we first parameterize a Heisenberg Hamiltonian in order to reproduce the experimental Curie temperature for a bulk system free of APBs, checking also the suitability of the parameters by calculating the corresponding exchange stiffness value using analytical and numerical methods (via simulating domain walls). Secondly we introduce APB defects in our system, allowing the new exchange interactions appearing across the interface being free parameters in order to study their influence in terms of their strength. We consider four different situations considering different exchange interactions across the anti-phase boundary. We first consider how the domain wall profile might be modified with the aim of determining for which cases the APB exchange is dominant rather than the bulk one. We will then consider the effects of applying a strong magnetic field in order to investigate how the anomalous saturation magnetization behaviour occurs. Finally, we study the stability of magnetic domain walls under thermal fluctuations in the presence of APBs to determine if an APB defect is a sufficient condition to have to two different magnetic domains.

We find that the presence of an APB defect prevents the magnetization from locally saturating when applying magnetic fields higher than the corresponding anisotropy field. Therefore, for samples grown with many APB defects, many regions of the material exist where the magnetization requires high magnetic fields to be saturated, explaining the anomalous saturation magnetization reported experimentally \cite{PhysRevLett.79.5162}. We also demonstrate that domain walls are stable under thermal fluctuations in the presence of APBs, explaining the high number of magnetic domains observed in samples with high density of APBs \cite{C7NR07143D}. Confirming APBs as  responsible for the different magnetic behaviours observed in magnetite thin films grown with different methods opens the door of tuning the magnetic properties of this system by growing samples with APB defects in a controlled way.

\section{Atomistic Spin Model of Magnetite}
To model a  magnetite thin film with APB defects we use a Heisenberg spin Hamiltonian, in which each magnetic moment is localized on its corresponding atom \cite{0953-8984-26-10-103202},
	
\begin{equation}
\mathcal{H}=-\frac{1}{2} \sum_{i\neq j} J_{ij}
\mathbf{S}_i \cdot \mathbf{S}_j - \sum_i k_{i,\mathrm{u}} S_{i,z}^2 - \sum_i \mu_i \mathbf{B} \cdot \mathbf{S}_i,
\label{Heis}
\end{equation}

where $\mathbf{S}_i$ is a unit vector describing the direction of the spin associated to the atom placed at site $i$, $S_{i,z}$ is its corresponding projection on the z, $J_{ij}$ is the exchange interaction between the spins $i$ and $j$, $k_{i,\mathrm{u}}$ represents an uniaxial energy term for the atom $i$, $\mu_i$ is the local atomic spin moment on each Fe site and $\mathbf{B}$ is the externally applied magnetic field.

Considering a cubic anisotropy term instead of an uniaxial one might lead to another type of domain wall different from the traversal one as described in \cite{6971574} for magnetite thin films. Therefore, as we wish to focus on the role that the exchange interactions across the APBs play on the anomalous saturation magnetization field and magnetic domain densities, we decided for simplicity to consider a uniaxial magnetic anisotropy term. The value we use for the anisotropy constant is $k_{i,\mathrm{u}}=3.26 \cdot 10^{-24}$ J/atom to ensure that we can fully contain a domain wall in our limited system size as discussed below.

The exchange interactions in magnetite are well known to be due to superexchange. Thus, their values strongly depend on the angle formed by the two interacting Fe atoms and the corresponding mediating O atom. In our model, the oxygen atoms are not simulated explicitly because they are non-magnetic and the exchange interactions are considered in the same way as direct exchange with an effective interaction of the usual Heisenberg form. For bulk magnetite, we fit the first nearest neighbor interactions in order to exactly reproduce the experimental N\'eel temperature, by considering a strong antiferromagnetic exchange interaction between the tetrahedral (Fe$_A$) and  octahedral (Fe$_B$) sublattices that reproduces the antiparallel orientation described by N\'eel \cite{refId0}, an antiferromagnetic exchange between tetrahedral Fe atoms and a ferromagnetic one between octahedral Fe atoms \cite{PhysRevB.47.5881}. The first nearest neighbors exchange interaction values for bulk magnetite obtained in this work are shown in Table \ref{Bulkexc}, together with their corresponding distances and bond angles.

\begin{table}[tb]
\caption{Exchange parameters used to reproduce bulk Magnetite $T_{\mathrm{C}}$ with their corresponding distances between Fe atoms and the bond angle formed with the oxygen atoms. }
\centering
\begin{tabular}{c c c c c c c c}
\hline\hline
& Exchange  &     & Energy  &   &Fe-Fe  & & Fe-O-Fe \\
&interaction & & ($10^{-22}$ J)& & Distance(A) & & bond angle\\
\hline
& $J_{\mathrm{Fe_A-O-Fe_A}}^{\mathrm{bulk}}$  &   &  $ -6.31 $  & &3.6 & & 80$^\circ$\\
& $J_{\mathrm{Fe_B-O-Fe_B}}^{\mathrm{bulk}}$  &   &  $ 9.76  $  & &2.9 & & 90$^\circ$\\
& $J_{\mathrm{Fe_A-O-Fe_B}}^{\mathrm{bulk}}$  &   &  $-50.7  $  & &3.5 & & 125$^\circ$\\
\hline
\hline\hline
\label{Bulkexc}
\end{tabular}
\end{table}

The time evolution of the spins is calculated via integration of the stochastic Landau-Lifshitz-Gilbert (sLLG) equation applied at the atomic level \cite{Ellis}, 

\begin{equation}
\frac{d \mathbf{S}_i}{dt} = - \frac{\gamma_i}{(1+\lambda_i^2)} \mathbf{S}_i \times [ \mathbf{B}_i + \lambda_i \mathbf{S}_i \times \mathbf{B}_i ],
\label{eq:llg}
\end{equation}

where $\lambda$, $\gamma_i = 1.76 \times 10^{11}$ T$^{-1}$s$^{-1}$ and $\mu_i$ are the damping constant, the gyromagnetic ratio and the magnetic moment associated to the spin $\mathbf{S}_i$ respectively. Here we assume a vanishing orbital magnetic moment giving local moments from Hund's rules of $\mu_{\mathrm{Fe_A}}= 5 \mu_B$ and $\mu_{\mathrm{Fe_B}} = 4.5 \mu_B$ for tetrahedral and octahedral sites respectively. The effective field acting on each spin, $\mathbf{B}_i=-\partial \mathcal{H}/\mu_i \partial \mathbf{S}_i + \boldsymbol{\zeta}_i,$ is calculated from the derivative of the atomistic Hamiltonian with respect to the spin $\mathbf{S}_i$ plus a stochastic magnetic field, $\boldsymbol{\zeta}_i$, which represents the thermal fluctuations of the spin system \cite{0953-8984-26-10-103202}. 




Although the damping constant has been predicted to have small values for magnetite \cite{4f9d0f6c8b614c83b6f2d46a108f9945}, we chose critical damping $\lambda = 1$ in this work in order to describe the influence of the APB on the quasi-equilibrium magnetic properties of magnetite. Therefore, we focus on the final magnetization state of our simulations but not on its dynamics. By increasing the damping constant we therefore decrease the computational time to reach equilibrium.



\section{Results}

\subsection{APB free magnetite system}
In order to study the properties of APBs in magnetite, we first create a rectangular stripe, free of defects,  with  length $L=118$ nm and a square cross section of $S \approx 1$ nm$^2$. In this geometry, the $x$ (elongated one), $y$ and $z$ directions correspond to the $(\overline{1}\overline{1}0)$, $(001)$ and $(\overline{1}10)$ crystal orientations respectively. For this system, we consider periodic boundary conditions (PBC) in the $y$ and $z$ directions and anti-periodic boundary conditions in the $x$ one. To consider anti-periodic boundary conditions (a-PBC) is the same  as considering ordinary PBC but changing the sign of all the exchange interactions in order to have an anti-parallel alignment between both sides of the system. The latter retains the bulk properties in our system at elevated temperatures, like the exchange stiffness ($A$) value while forcing our magnetite stripe to have two magnetic domains and therefore to have a domain wall whose width can be determined. 

In this geometry, free of APB defects, a domain wall at temperature $T=0K$ is fully contained  because the  macroscopic anisotropy energy and the exchange stiffness, corresponding to our atomistic parameters, are $K=1.36\cdot 10^5$ J/m$^3$ and $A=1.75\cdot 10^{-11}$ J/m respectively with a domain wall width of $\delta_{DW} = \pi \sqrt{A/K} = 35.63$ nm. It should be pointed out that, although the atomistic exchange parameters have been fitted to reproduce realistic magnetic properties of magnetite ($T_N$), the anisotropy energy value has been chosen to be large with the aim of reducing the system size needed in our simulations. Therefore, the free domain wall width calculated here should be smaller than the real one \cite{6971574}.  

To generate equilibrated domain wall profiles we initialise half of the net ferrimagnetic magnetization along the $+x$ direction and the other half along $-x$ direction, forming a complete $180^\circ$ domain wall with zero width. We then let the system evolve for $1$ ns, which allows enough time to form an equilibrium domain wall whose center is placed in the middle of the system. Analytically the equivalence between the macroscopic anisotropy ($K$) and the atomistic one ($k_i^u$) has been obtained by considering the \textit{magnetic} atomic volume as $V_{\mathrm{mag,at}}=a^3/N_{\mathrm{mag,at}}= 0.0242$ nm$^3$, where $a=0.834$ nm is the lattice parameter for magnetite, $a^3$ is the volume of the unit cell and $N_{\mathrm{mag,at}}=24$ is the number of magnetic atoms per unit cell. To calculate the exchange stiffness value ($A$) from our atomistic parameters we have extended the description done by Aharoni \cite{aharoniBOOK96} for a ferromagnet to a ferrimagnet. To do this, we have considered small variations around $180^\circ$, instead of $0^\circ$, when the exchange interaction corresponds to spins belonging to different sublattices because they are strongly antiferromagnetically coupled. The equation we obtained is similar to the one published in \cite{MorenoPRB2016},

\begin{equation}
    A^\nu=\frac{1}{4V} \left( \sum_{i j }^{id(i)=id(j)}  J_{ij}\left(r^\nu_i -r^\nu_j \right)^2 - \sum_{i j }^{id(i)\neq id(j)} J_{ij}\left(r^\nu_i -r^\nu_j \right)^2\right),
\label{Aafm}
\end{equation}

where $V$ is volume of the system, $i$ and $j$ run over all atoms, $v = x,y,z$ are  vector coordinates, $J_{ij}$ are the exchange interactions, $r^\nu_i$ is the $\nu$ component of the position of the atom $i$, $id(i)$ and $id(j)$ are the sublattices to which the atoms $i$ and $j$ correspond respectively.
Its calculated value, apart from being isotropic, is in good agreement with values found in literature \cite{C7NR07143D,HEIDER1287}, giving confidence in the micromagnetic exchange parameterization. We note that the above expression is only useful for computing the domain wall width in the case of uniaxial anisotropy. Domain walls in cubic systems are more omplex and will be the suject of future work.
\begin{figure}[!tb]
\centering
\includegraphics[width=1.0\columnwidth ,angle=0]{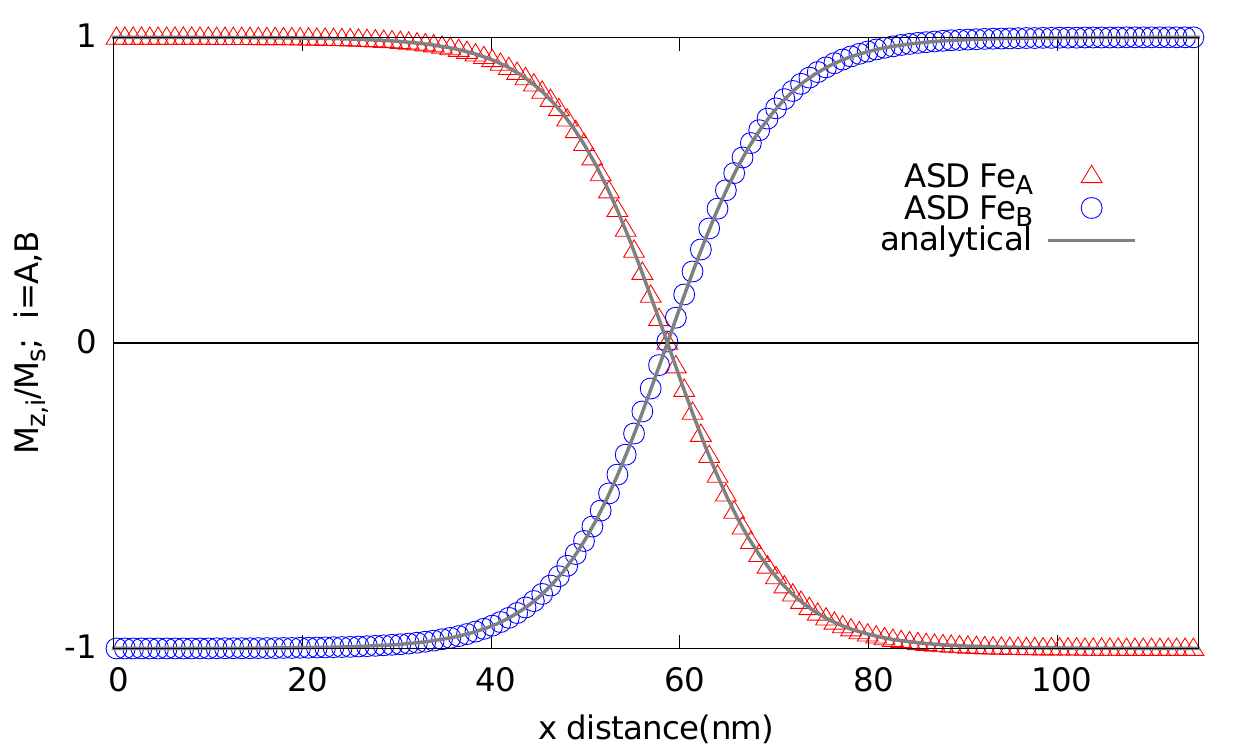}
\caption{Magnetic configuration of a domain wall for a magnetite stripe without APB defects at $T=0$ K. Red triangles and blue circles represent the numerical results for the  z component of the magnetization for the tetrahedral and the octahedral sublattices respectively. The  grey line represents the analytical domain wall profile obtained from the macroscopic anisotropy and exchange stiffness ($m_z= \mathrm{tanh}((x-x_0)\pi /\delta_{DW}) $).} 
\label{BulkDw}
\end{figure}

In Fig.~\ref{BulkDw} we show the simulated sublattice resolved domain wall together with the analytical profile, $m_z= \mathrm{tanh}((x-x_0)\pi /\delta_{DW}) $, showing a perfect match between both analytical and numerical results.

\subsection{Anti-phase boundary properties}

\begin{figure}[!tb]
\centering
\includegraphics[width=1.0\columnwidth ,angle=0]{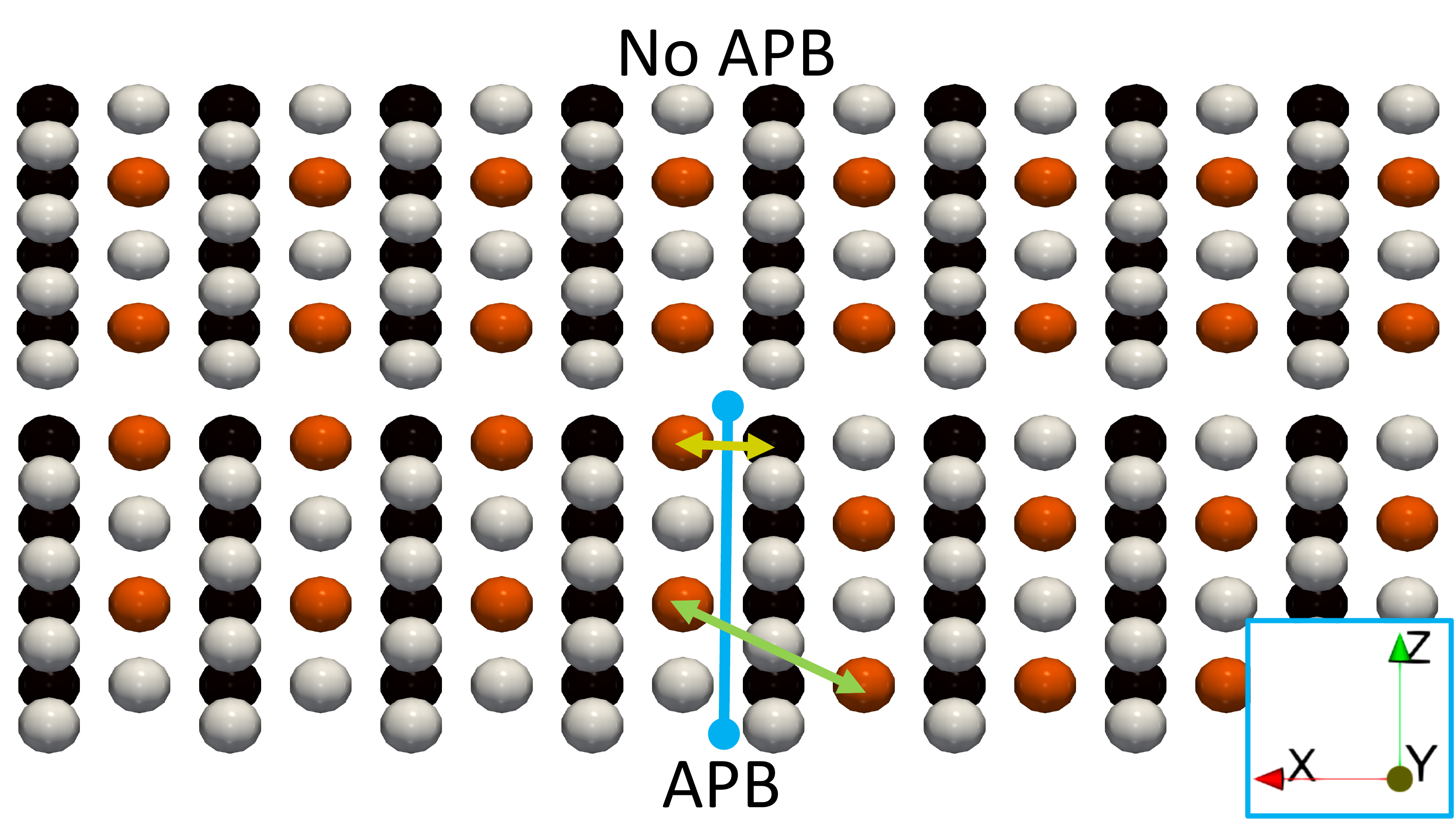}
\caption{Representation of an APB defect in magnetite created by a displacement $a_0\sqrt{2}/4$ on the z direction. Top figure shows the bulk system free of APB, in the bottom one an APB defect has been introduced (blue line). X,Y and Z represent the $(\overline{1}\overline{1}0)$, $(001)$ and $(\overline{1}10)$  orientations respectively. Black and  orange  spheres represent the octahedral and tetrahedral Fe atoms respectively, white one are the oxygen atoms. Yellow and green arrows indicate the new $J_{AB,new}$ and $J_{BB,new}$ exchange interactions across the interface. } 
\label{APBdefect}
\end{figure}

To introduce an APB defect in our geometry, we divide the system in two halves in the $x$ direction and displace one them $a_0\sqrt{2}/4$ along the $z$-axis ($\overline{1}10$), shown schematically in Fig.~\ref{APBdefect}, in comparison with the magnetite stripe without the APB defect.

Due to this dislocation, the atomic distances as well as the angles formed by the interacting Fe-O-Fe sets are modified across the interface. Consequently, the number of bulk exchange interactions in the APB is reduced drastically but some interactions still remain. Additionally, new exchange interactions are created with different distances between iron atoms and bond angles with the oxygen one\cite{PhysRevLett.79.5162}. In Fig.~\ref{APBdefect}, these new exchange interactions are schematically represented with arrows and  their distances, bond angles and the number of them per unit cell are presented together with the corresponding bulk values in Tab.~\ref{APBexc}.



\begin{table}[!tb]
\caption{Characterization of the  exchange interactions appearing across the APB with their corresponding distances between $Fe$ atoms, the bond angle formed with the oxygen atoms and the number of them per unit cell across the interface. Subindex $a$ points the values taken from literature \cite{PhysRevLett.79.5162}.}
\centering
\begin{tabular}{c c  c c c c c c c }
\hline\hline
& Exchange  &        &   &Fe-Fe  & & Fe-O-Fe & & Number of  \\
&interaction & &  & Distance(A) & & bond angle & & interactions\\
\hline
& $J_{\mathrm{Fe_A-O-Fe_B}}^{\mathrm{APB}}$  &     $  $  &  &1.8 & & 55$_a^\circ$ & & 4\\
& $J_{\mathrm{Fe_B-O-Fe_B}}^{\mathrm{APB}}$  &     $   $  & &4.16 & & 180$^\circ$    & & 8\\
& $J_{\mathrm{Fe_B-O-Fe_B}}^{\mathrm{Bulk}}$  &     $   $  & &2.9 & & 90$^\circ$   & & 8\\
& $J_{\mathrm{Fe_A-O-Fe_B}}^{\mathrm{Bulk}}$  &     $  $  &  &3.5 & & 125$_a^\circ$ & & 16\\

\hline
\hline\hline
\label{APBexc}
\end{tabular}
\end{table}

The distances, bond angles and number of interactions we found across the APB match with the previously described in \cite{PhysRevLett.79.5162} (APB type 3), in which the closest interaction, $J_{\mathrm{Fe_A-O-Fe_B}}^{\mathrm{APB}}$, is proposed to be the main interaction responsible for the anomalous magnetic behaviour in magnetite thin films. Nevertheless, in this work, for completeness, we also consider the effect of the second nearest neighbour $J_{\mathrm{Fe_B-O-Fe_B}}^{\mathrm{APB}}$ to explain the atypical magnetic properties. 

Since it is not possible to fit the value of the APB exchange interactions to reproduce any experimental result as we did for the bulk, they are considered as free parameters. However, some considerations can be taken into account in order to restrict their value range. On the one hand, the distance between Fe atoms in $J_{\mathrm{Fe_A-O-Fe_B}}^{\mathrm{APB}}$ is small enough to consider that this exchange interaction might be direct, not mediated by any oxygen atoms. Therefore, due to this distance is in between the first nearest neighbor and the second one for bulk iron we could expect $J_{\mathrm{Fe_A-O-Fe_B}}^{\mathrm{APB}}$ to be positive and relatively large \cite{PhysRevB.64.174402}.
On the other hand, the distance between Fe atoms for $J_{\mathrm{Fe_B-O-Fe_B}}^{\mathrm{APB}}$ is big enough to consider that the contribution from direct exchange must be small, however, as the angle they form with the oxygen atom is $180^{\circ}$, we expect this interaction to be antiferromagnetic and strong \cite{SAWATZKY197637}. 

\begin{figure*}[!tbp]
\centering
\includegraphics[width=2.0\columnwidth ,angle=0]{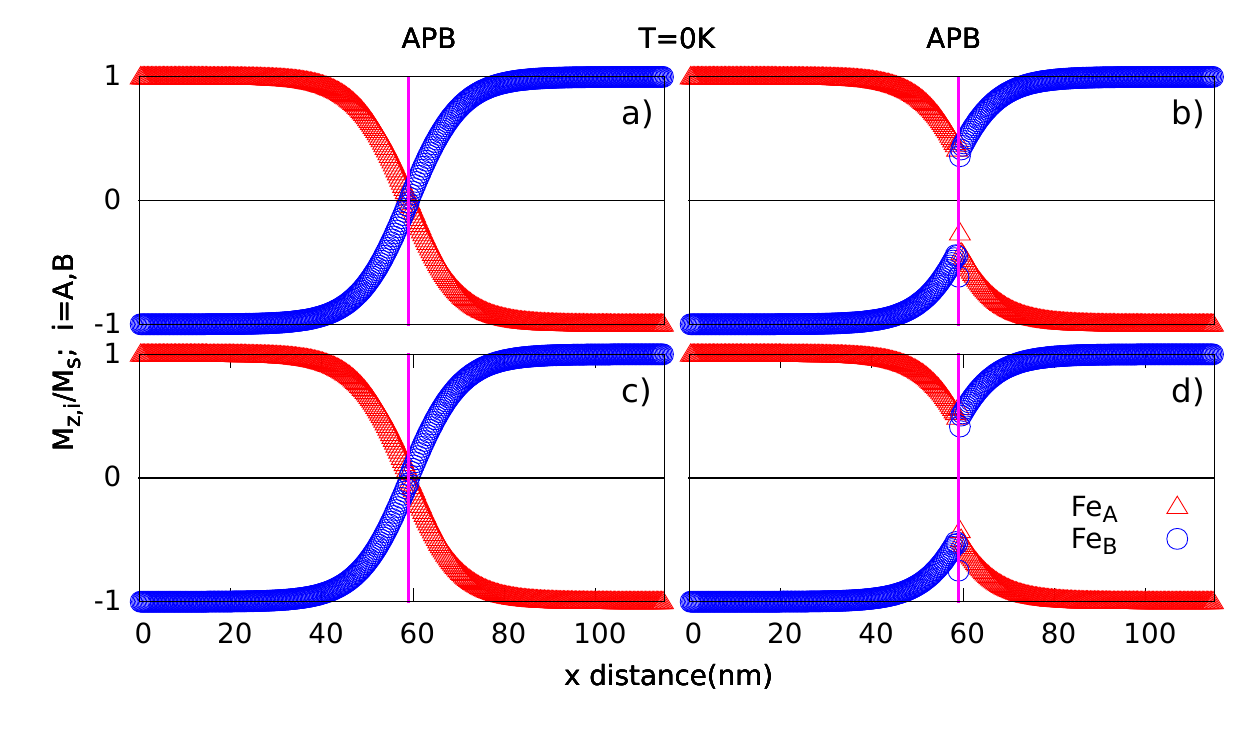}
\caption{Sublattice resolved domain wall in a magnetite stripe with an APB defect in the middle, using different parameterizations of the exchange interactions across the APB. Red triangles and blue circles represent the z component of the magnetization for the tetrahedral and the octahedral sublattices respectively. The pink line represents the position of the APB. Figures a), b), c) and d) correspond to case 1, 2, 3 and 4 respectively (see Table \ref{cases}).}
\label{DwAPB}
\end{figure*}

For both APB exchange interactions taken into account, we consider a wide range of possible values, ranging from 0 to  $J_{\mathrm{Fe_A-O-Fe_B}}^{\mathrm{APB}}= 5.07\cdot 10^{-21} J$ and $J_{\mathrm{Fe_B-O-Fe_B}}^{\mathrm{APB}}=-5.86\cdot 10^{-21} J$ respectively. For each parameterization, we simulate a domain wall at temperature $T=0K$ in order to determine the APB influence on the domain wall profile.  In Fig \ref{DwAPB}, we show the simulated domain walls for the four extreme parameterization cases, which are described in Tab.~\ref{cases}.  

\begin{table}[tb]
\caption{Extreme parameterization cases displayed in this work}
\centering
\begin{tabular}{c c  c c c c c c c }
\hline\hline
& Case  &        &   &$J_{\mathrm{Fe_A-O-Fe_B}}^{\mathrm{APB}}$  & & $J_{\mathrm{Fe_B-O-Fe_B}}^{\mathrm{APB}}$ & &  \\
\hline
\hline
& 1  &     $  $  &  &0.0 & & 0.0 & & \\
& 2  &     $   $  & &0.0 & &$-5.86\cdot 10^{-21} J$     & & \\
& 3  &     $   $  & &$5.07\cdot 10^{-21} J$ & & 0.0  & & \\
& 4  &     $  $  &  &$5.07\cdot 10^{-21} J$ & & $-5.86\cdot 10^{-21} J$ & & \\

\hline
\hline\hline
\label{cases}
\end{tabular}
\end{table}

From Fig.~\ref{DwAPB} we observe how the domain wall profile is clearly modified by the presence of an APB defect. For the case of neglecting the new exchange interactions (case 1) the $\tanh(x)$ profile is not modified. However, as the exchange interactions across the APB increase,  $m_z$ starts to be a discontinuous function of the distance.
It is observed, comparing the cases of considering only one of the new exchange interactions and neglecting the other one (cases 2 and 3), that the effect produced by  $J_{\mathrm{Fe_B-O-Fe_B}}^{\mathrm{APB}}$ is stronger than the one produced by $J_{\mathrm{Fe_A-O-Fe_B}}^{\mathrm{APB}}$, in contrast with what was suggested in \cite{PhysRevLett.79.5162}. This is due to the number of  $J_{\mathrm{Fe_B-O-Fe_B}}^{\mathrm{APB}}$ bonds compared to the number of  $J_{\mathrm{Fe_B-O-Fe_B}}^{\mathrm{bulk}}$ exchange interactions across the interface is the same (see Table \ref{APBexc}), therefore, the antiferromagnetic exchange dominates rather than the bulk ferromagnetic exchange, once its value is larger. Thus, as the value for $J_{\mathrm{Fe_B-O-Fe_B}}^{\mathrm{bulk}}$ is small (see Table \ref{Bulkexc}), we can consider that values for $J_{\mathrm{Fe_B-O-Fe_B}}^{\mathrm{APB}}$ larger than $J_{\mathrm{Fe_B-O-Fe_B}}^{\mathrm{bulk}}$ are feasible.  For the case of $J_{\mathrm{Fe_A-O-Fe_B}}^{\mathrm{APB}}$, the number of interactions is $4$ times lower than for $J_{\mathrm{Fe_A-O-Fe_B}}^{\mathrm{bulk}}$. The bulk antiferromagnetic exchange between sublattices in magnetite is particularly strong and so a value for $J_{\mathrm{Fe_A-O-Fe_B}}^{\mathrm{APB}}$ which is $4$ times larger than $J_{\mathrm{Fe_A-O-Fe_B}}^{\mathrm{bulk}}$ will be excessively large and unrealistic.

\subsection{Saturation of anti-phase boundary defects}
\begin{figure}[!tb]
\centering

\includegraphics[width=1.0\columnwidth ,angle=0]{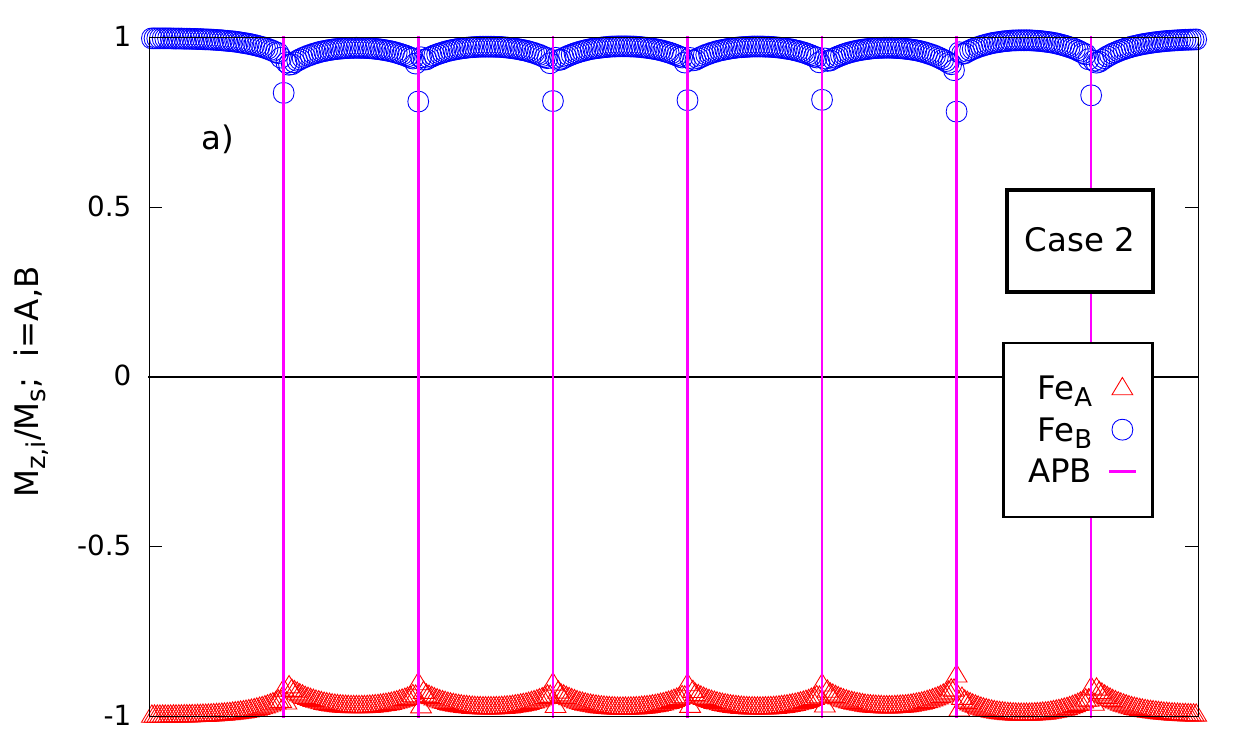}
\includegraphics[width=1.0\columnwidth ,angle=0]{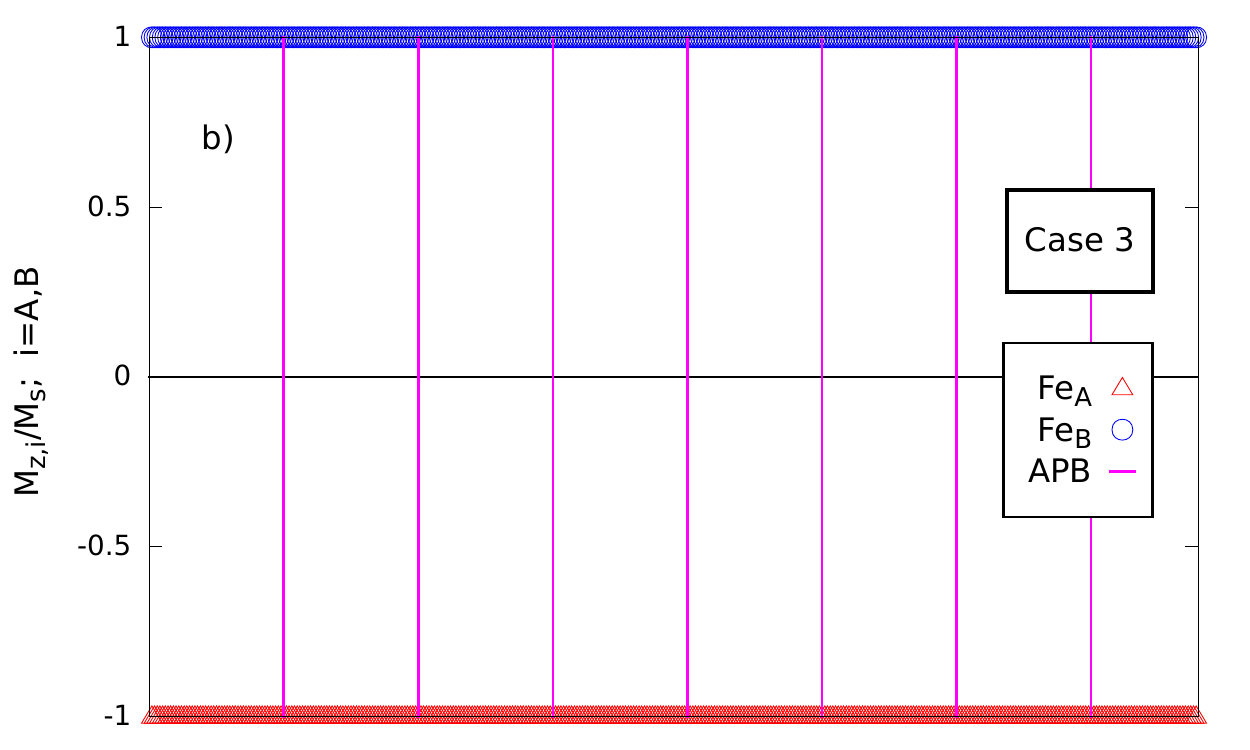}
\includegraphics[width=1.0\columnwidth ,angle=0]{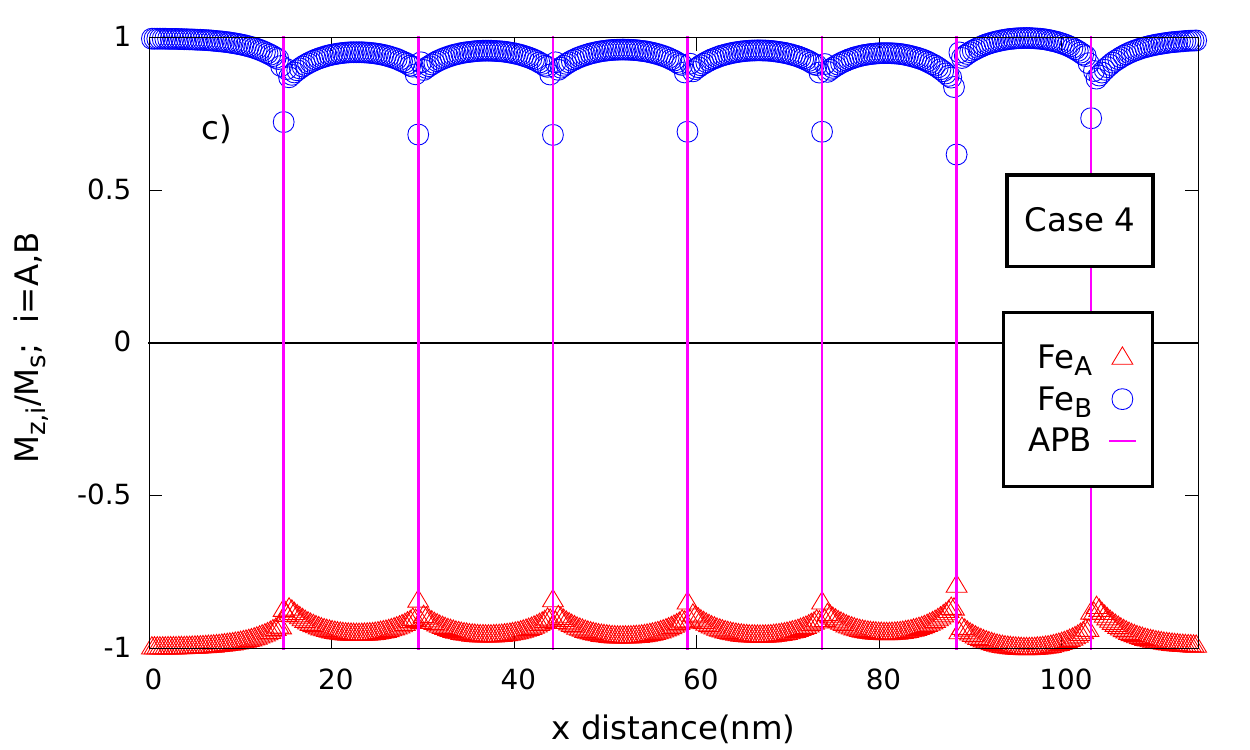}
\caption{Sublattice resolved magnetization state in a magnetite stripe with 7 APB defects (pink lines)  when a magnetic field is applied in the z direction. Red triangles and blue circles correspond to tetrahedral and octahedral sublattices respectively. Figures a), b) and c) correspond to the cases 2,3 and 4 respectively (see Table \ref{cases}).} 
\label{APBwithH}
\end{figure}

Using  the domain wall spin configurations obtained for each APB parameterization as an initial state, we restart our simulations in order to investigate how the APB exchange interactions could influence the anomalous saturation magnetic field for magnetite thin films. With this aim, the anti-PBC on the $x$ edges are removed to let the system saturate under the influence of a magnetic field $\mathbf{B}$. The magnetization state after applying a magnetic field of 1 T, during $1$ ns, in the $z$ direction, indicates that  the applied magnetic field is not able to saturate the magnetization where the APB defect is placed if $J_{\mathrm{Fe_B-O-Fe_B}}^{\mathrm{APB}}$ is considered. At the APB, the magnetization reduction observed for the parameterization case 4 is 20 $\%$ for the $Fe_A$ and 40 $\%$ for the $Fe_B$ sublattices. Note that this field strengh is larger than the anisotropy field ($\mu_0 H_K \approx 0.5 T$).  To confirm this result is not an artifact arising from using the domain wall as an initial configuration, we create the same system but considering 7 uniformly distributed APB defects and, as an initial condition, a random spin configuration. In Fig.~\ref{APBwithH}, we show the final magnetic state after applying the same magnetic field strength applied for 1 ns. The results confirm that the magnetization cannot be saturated at the APBs if only $J_{\mathrm{Fe_B-O-Fe_B}}^{\mathrm{APB}}$ is considered. However, by comparing the cases 2 and 4, we can assert that the $J_{\mathrm{Fe_A-O-Fe_B}}^{\mathrm{APB}}$ presence increases this effect. Moreover, the magnetization  barely saturates in between two APB due to the short distance between them, suggesting that  the number density of defects as a crucial parameter responsible for the magnetization reduction at high magnetic fields.

\begin{figure}[!tb]
\centering
\includegraphics[width=1.0\columnwidth, angle=0]{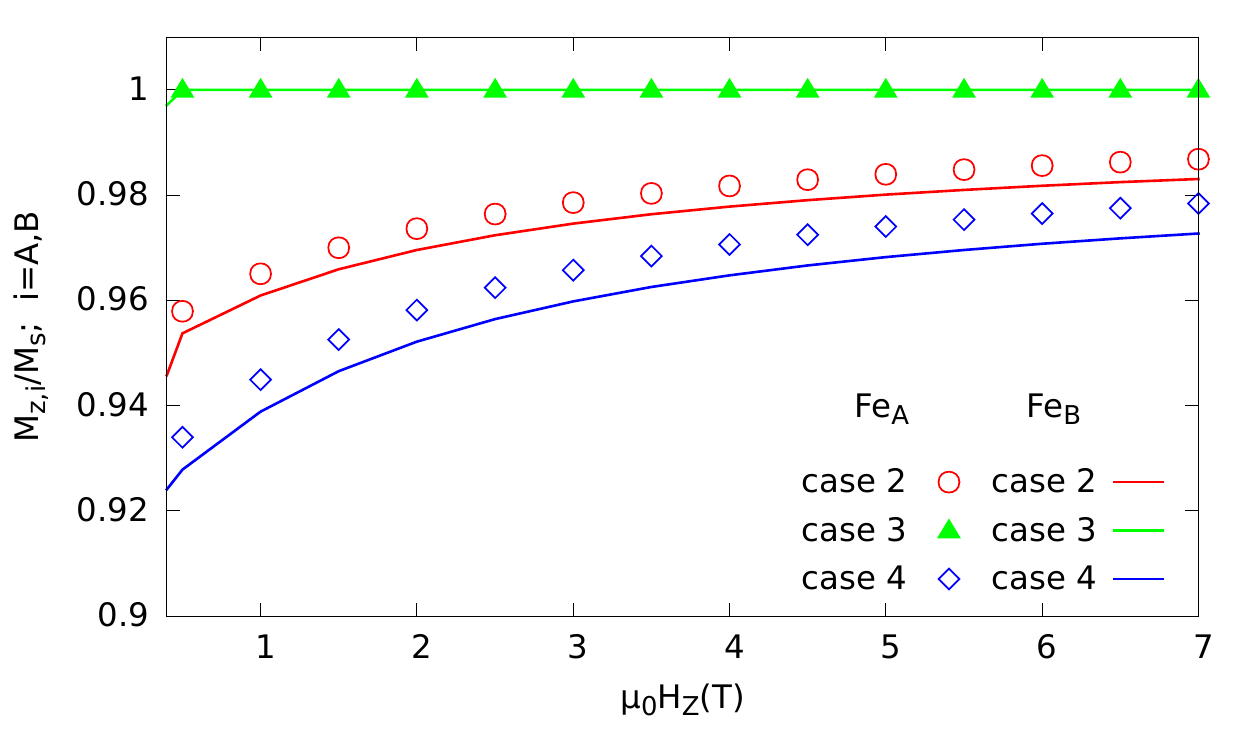}
\caption{Partial hysteresis loop of a magnetite stripe with 7 APB defects for the different parameterization cases displayed in table \ref{cases}. Points and lines represent the spatial average magnetization for the octahedral and tetrahedral sublattices respectively.} 
\label{Hyst}
\end{figure}

As the new exchange interactions appearing across the APB are responsible for the non-saturating regions, the magnetic field needed to saturate the sample is likely to be very high, as reported experimentally \cite{PhysRevLett.79.5162}. Note that the exchange contribution for the effective field acting on Eq.~\ref{eq:llg} usually is a tens of tesla in terms of order of magnitude and so likely to dominate the magnetic coupling at the APB interface. To confirm this fact, we have increased the magnetic field in steps of $0.5T$, calculating the mean equilibrium sublattice resolved magnetization for each case, until a maximum of $7T$, which is the experimental value addressed in \cite{PhysRevLett.79.5162}. Results are displayed in Fig.~\ref{Hyst}, showing that even for the highest magnetic field the magnetization is not saturated. 

These results indicate that the main interactions responsible of the anomalous saturation field are the $J_{\mathrm{Fe_B-O-Fe_B}}^{\mathrm{APB}}$ interactions but the $J_{\mathrm{Fe_A-O-Fe_B}}^{\mathrm{APB}}$ energy increases this effect. It should be noted that the real density of APBs on magnetite thin films usually is greater than the one we are considering \cite{McKenna2014}, thus, the effect of the notch observed, as well as the non saturating region between APBs, will be more pronounced in the total magnetization measurements.  

\subsection{Temperature dependence of anti-phase boundary defects}

To investigate why the number of magnetic domains observed in magnetite thin films is proportional to the volume fraction of APBs in the sample \cite{C7NR07143D}, we enable thermal fluctuations in our calculations  to determine the stability of a multi-domain state. Specifically, we consider two different temperatures, $T=1$K and room temperature $T = 300$K. The former temperature is considered because it has a  weak contribution to the effective field in eq. \ref{eq:llg}, allowing us to understand the role that each of the new exchange interactions is  playing on the domain stability. For this case, we work with the same geometry as before with a single APB defect. For the $T=300$ K case, we consider a wider system because, for higher temperatures, a bigger spatial average is required to obtain a smooth magnetization profile. Explicitly, the new cross sectional area under consideration is $S \approx 36$ nm$^2$ but the length is kept same as before. Both systems present a single APB defect  placed in the middle of the system.

\begin{figure}[!tb]
\centering
\includegraphics[width=1.0\columnwidth ,angle=0]{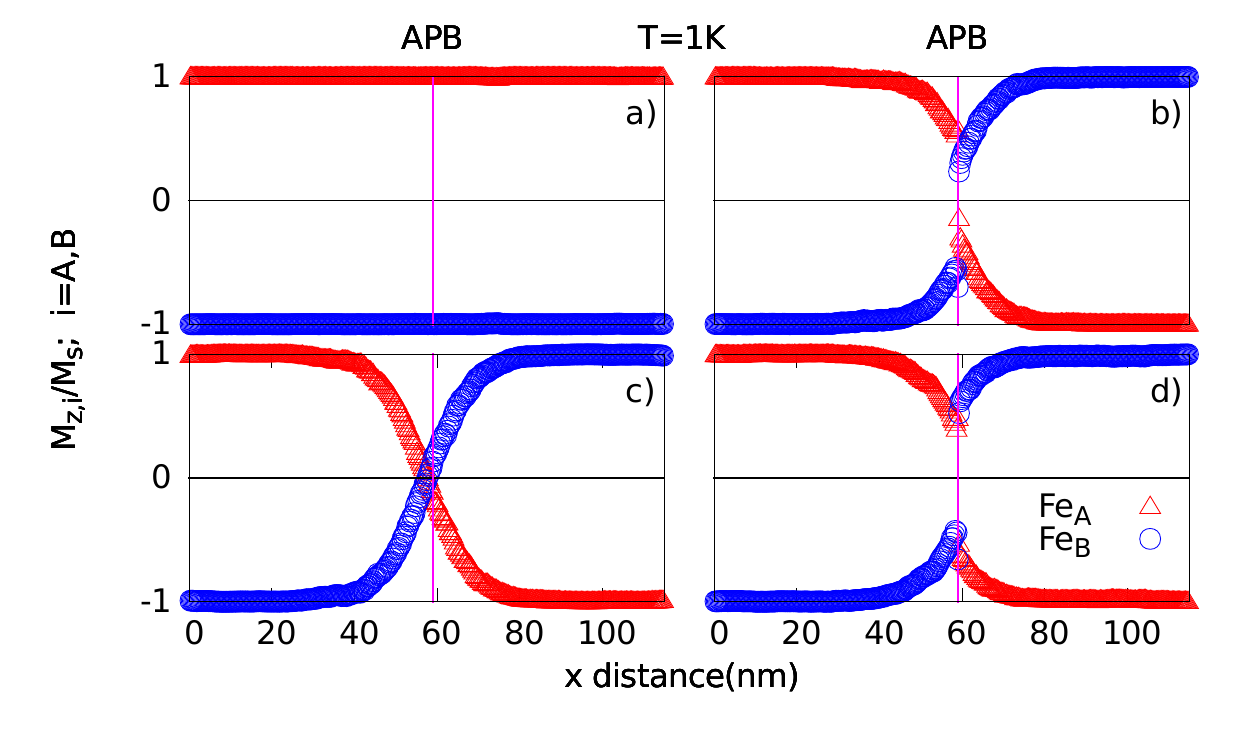}
\includegraphics[width=1.0\columnwidth ,angle=0]{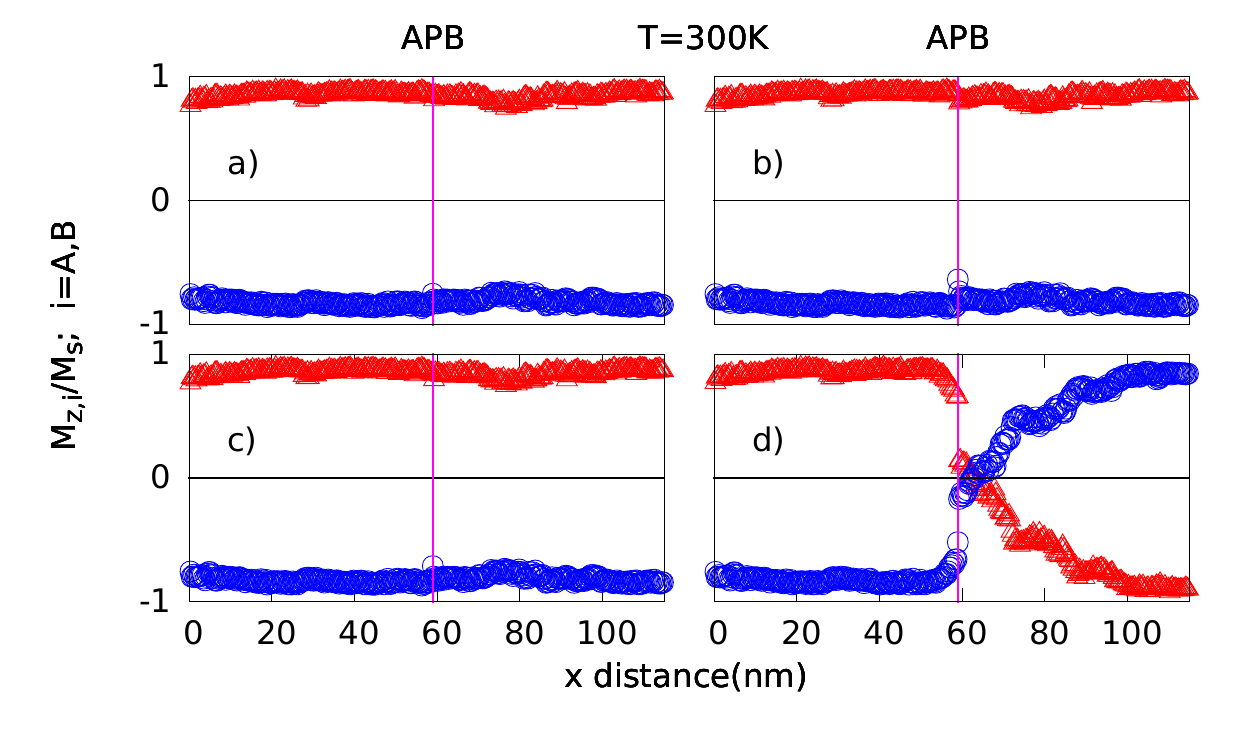}
\caption{The sublattice resolved magnetization state after $50ns$ simulation of a domain wall under the effect of a thermal field in the presence of an APB defect for two different temperatures. $T=1K$ and $T=300K$ figures are presented above and below respectively. Red triangles and blue circles represent the z component of the magnetization for the tetrahedral and the octahedral sublattices respectively. The pink line represent the position of the APB.  Figures a), b), c) and d) correspond to case 1, 2, 3 and 4 respectively (see Table \ref{cases})  } 
\label{DwAPBT}
\end{figure}

For both cases, the initial condition set up for the magnetization consists of two magnetic domains separated by the APB, aligned antiparallel and pointing in the anisotropy direction. The main difference with previous domain wall calculations, apart from including temperature, is done by removing the a-PBC conditions. Thus, the two introduced magnetic domains are not forced to remain in the system and one them will be erased by the effect of thermal fluctuations if the APB does not prevent it. We let the system to evolve for $50ns$, which is sufficient time for the domain wall to escape from the system. The final magnetic configurations for the extrema of the parameterization cases (Table \ref{cases}) are displayed in Fig.~\ref{DwAPBT}.

From Fig. \ref{DwAPBT} we observe that in the absence of APB exchange interactions (case 1), the thermal fluctuations at both temperatures are sufficient to remove the two-domain state previously introduced. However, as larger exchange interactions are included, the domain wall starts to get pinned and a multidomain state is stable for small thermal fluctuations ($T=1$K). In contrast with the previous results at $T = 0$K, a domain wall under the effect of thermal fluctuations is also stable if we neglect $J_{\mathrm{Fe_B-O-Fe_B}}^{\mathrm{APB}}$ but consider $J_{\mathrm{Fe_A-O-Fe_B}}^{\mathrm{APB}}$. This effect is also presented for higher temperatures ($T=300K$), however, as the thermal fluctuations become stronger, higher values for the APB exchange interactions are required to pin the domain wall at the APB. In fact, only for the cases in which both exchange interactions are considered and their values are high does the domain wall remain in the system after $50ns$. It might be possible that, even for parameterization case 4, the two domain state is removed if considering longer simulation. Nevertheless, results for both temperatures have shown than when increasing the value of the exchange interactions the stability of the two domain state increases too. Hence, to stabilize a two domain state for 300K requires higher exchange interactions, which can be still considered realistic. Therefore, we can confirm APBs as source of magnetic domains, with magnetic domain walls pinned at APBs, only due to the presence of new exchange interactions across it. A higher density of APB defects is also likely to increase the thermal stability of the multidomain state.

\section{Conclusions}
In this work we have modelled a magnetite system with and without APB defects using atomistic spin dynamics, focusing our analysis on the role that the  exchange interactions play across the APB interface. For the bulk case, we have fitted the first nearest neighbors exchange interactions in order to properly reproduce the experimental Curie temperature. Additionally, we obtain numerically and theoretically an exchange stiffness value which is in good agreement with the values previously published. After introducing an APB defect in the system by displacing half of the system in the $(\overline{1}10)$ direction, we find that the number of new exchange interactions in the APB and their corresponding distances and angles match with those previously reported in \cite{PhysRevLett.79.5162}.

For this defect, we consider a wide range of possible values for the new exchange interactions  $J_{\mathrm{Fe_A-O-Fe_B}}^{\mathrm{APB}}$ and $J_{\mathrm{Fe_B-O-Fe_B}}^{\mathrm{APB}}$, nevertheless,  with the aim of describing  qualitatively their corresponding  effect on the magnetization,  we  focus the results of this work on four extreme parametrization cases described in Table \ref{cases}. We show that, although  the $J_{\mathrm{Fe_A-O-Fe_B}}^{\mathrm{APB}}$ interactions were suggested to be the main ones responsible for the anomalous magnetic behaviour in magnetite thin films, it is likely not to be the case. Both the  anomalous saturation field and the high density of magnetic domains could be explained in terms of $J_{\mathrm{Fe_B-O-Fe_B}}^{\mathrm{APB}}$. This is due to the fact that the number of $J_{\mathrm{Fe_B-O-Fe_B}}^{\mathrm{APB}}$ bonds across the interface is the same as the $J_{\mathrm{Fe_B-O-Fe_B}}^{\mathrm{Bulk}}$, while the number of $J_{\mathrm{Fe_A-O-Fe_B}}^{\mathrm{APB}}$ is lower than $J_{\mathrm{Fe_A-O-Fe_B}}^{\mathrm{Bulk}}$.

On the one hand we show that, for the case of the saturation field, locally the magnetization cannot be saturated in the APB if $J_{\mathrm{Fe_B-O-Fe_B}}^{\mathrm{APB}}$ is considered. This effect comes up due to the antiferromagnetic exchange of $J_{\mathrm{Fe_B-O-Fe_B}}^{\mathrm{APB}}$ and it could be increased by considering a ferromagnetic exchange in $J_{\mathrm{Fe_A-O-Fe_B}}^{\mathrm{APB}}$. On the other hand, both exchange interactions produce pinning effects on the domain walls, demonstrating that a multidomain state is stable under the effect of thermal fluctuations due to the presence of APBs on the system. Because of when taking into account the new exchange interactions arising from the APB defect we  reproduce the high saturation field as well as the stability of the magnetic domains observed in magnetite thin films, we confirm them as the responsible for the anomalous magnetic properties observed experimentally.

\section*{acknowledgments}
The financial support of the Engineering and Physical Sciences Research Council (Grant No. EPSRC EP/P022006/1) is gratefully acknowledged. We gratefully acknowledge the provision of computer time made available on the \textsc{viking} cluster, a high performance compute facility provided by the University of York. 

\bibliography{sample}

\end{document}